\documentclass[a4paper,english,11pt]{article}
\usepackage{graphicx}
\usepackage{float}
\usepackage{amsmath} 
\usepackage{amssymb}
\usepackage{amsfonts} 

\usepackage{graphics}
\usepackage{eepic,epsfig}
\usepackage{cite}

\newcommand {\OMIT}[1]{}

\begin{document}

\begin{center}

{ \Large {Joint segmentation of many aCGH profiles using fast group LARS}}

\vspace{0.3cm}

Kevin Bleakley\,$^{\rm a, b, c}$ and Jean-Philippe Vert\,$^{\rm a, b, c}$

\vspace{0.3cm}

{$^{\rm a}$Mines ParisTech, Centre for Computational Biology,
35 rue Saint-Honor\'e,
F-77305 Fontainebleau Cedex, France,
$^{\rm b}$ Institut Curie, F-75248, Paris, France and $^{\rm c}$ INSERM, U900, F-75248, Paris, France}
\end{center}

\begin{center}
kevbleakley@gmail.com, \,\, Jean-Philippe.Vert@mines-paristech.fr 
\end{center}
\vspace{0cm}

\begin{abstract}

\begin{sloppypar}
Array-Based Comparative Genomic Hybridization (aCGH)  is a  
method used to search for genomic regions with  copy numbers variations. 
For a given aCGH profile, one challenge is to accurately segment it into regions of constant copy number.
Subjects sharing the same disease status, for example a type of cancer, often have aCGH profiles with
similar copy number variations, due to duplications and deletions relevant to that particular disease. 

We introduce a constrained optimization algorithm that jointly segments aCGH profiles of many subjects. 
It simultaneously penalizes the amount of freedom the set of profiles have to jump from one level of constant copy number to another,
at genomic locations known as breakpoints.
We show that breakpoints shared by many different profiles tend to be found first by the algorithm, even in the presence of significant
amounts of noise. 

The algorithm  can be formulated as a group LARS problem. We propose an extremely fast way to find the solution path,
i.e., a sequence of shared breakpoints in order of importance.
For no extra cost the algorithm smoothes all of the  aCGH profiles  into piecewise-constant regions of equal copy number, giving low-dimensional 
versions of the
original data. These can  be shown for all profiles on a single graph, allowing for intuitive visual interpretation.
Simulations and an implementation of the algorithm on bladder cancer aCGH profiles are provided. 
\end{sloppypar}

\end{abstract}


\section{Introduction} \label{intro}

Array-based Comparative Genomic Hybridization (aCGH) is a technique that aims to detect chromosomal aberrations on a genomic scale in a single
experiment. In tumors for example, chromosomal aberrations include  tumor suppressor genes being inactivated by deletion, and oncogenes being activated by
duplication. Such changes to the underlying ``normal'' genomic state, known as copy number variations (CNVs) provide information related to the current disease state of the individual. 

Many cancers are known to exhibit recurrent CNVs in specific locations in the genome, for example monoploidy of chromosome 3 in uveal melanoma \cite{Speicher}, 
loss of chromosome 9  in bladder carcinomas \cite{Blaveri2005Bladder}, loss of 1p and gain of 17q in
neuroblastoma \cite{Bown,VanRoy}
and amplifications of 1q, 8q24, 11q13, 17q21-q23 and 20q13 in breast cancers \cite{Yao}.  

aCGH allows rapid mapping of CNVs of a tumor sample at the genomic level \cite{Pinkel}.  Originally, the technique was based on
arrays with several thousand large insert clones (e.g., bacterial artificial chromosomes or BACs) with a megabase resolution. More recently, this has
been improved using oligonucleotide-based arrays with up to hundreds of thousands of probes,  bringing the resolution down to several kilobases
\cite{Gershon}. 
Each spot on an aCGH array contains  amplified or synthesized DNA of a particular region of the genome. The array is hybridized with DNA extracted from a sample of interest, and in most cases with
(healthy) reference DNA. Each of the two samples is then labeled with its own fluorochrome and the ratio of fluorescence between the 
two samples is expected to reveal the ratio of DNA copy number at each position of the genome. Usually the logarithm of this ratio is then taken and the values (profile) plotted
linearly along the genome, as shown for example in Fig. \ref{piecewise}(b)-(c). 
%
%
%
%



One challenge related to aCGH profiles is to detect regions of constant copy number, separated by breakpoints, in the presence of
significant amounts of noise. Many groups have attempted to answer this question when treating a single profile, with various methods for segmentation and smoothing of 1-dimensional profiles \cite{Olshen2004Circular,Hupe, Picard2005,Fridlyand,Wang2005method,Picard2007,Huang2007Robust,Tibshirani2008}.
Recently, approaches have been suggested for dealing with \emph{multiple} aCGH profiles. 
Indeed, we often have experimental data from a series of patients who have the same disease status, or several 
groups of patients, each with a particular disease status. The goal of jointly treating several profiles is to extract breakpoints and copy number variations that are globally representative of the disease status of those patients.
The STAC algorithm \cite{Diskin} uses a novel search strategy
to find regions of copy number gains and losses that are statistically significant across a set of profiles.
It assigns $p$-values  to each location on the genome using a 
multiple testing corrected permutation approach. The approach of \cite{Klijn} is based on kernel regression.
One disadvantage of this is that results are dependent on
the width of the kernel window, so 
a region may exhibit a statistically significant copy number variation for one width, but not for another, leading to difficulties
in interpretation. The approach in \cite{Rouveirol} 
is to first discretize  profiles independently into gains, losses and normal. Then, they look for regions in which a significant number of the
profiles share gains or losses using a Boolean framework and theoretical results for pattern searching. 
This method potentially loses pertinent information in the discretization step. 
Finally, the approach in \cite{Picard2007b} is to simultaneously segment a set of profiles using mixed linear models to account for both
covariates and correlations between probes.  They proposed an EM-type algorithm that uses dynamic programming during the
segmentation step. Though the computational cost was not provided, the algorithm appears computationally
intensive. 

In this article, we present an algorithm that  jointly renders a set of $n$ profiles piecewise-constant, with the
intended goal of uncovering a set of breakpoints and regions of constant copy number that
are pertinent with respect to the whole set of profiles.  
%
%
This algorithm generalizes a
Fused Lasso-type algorithm that was used to find breakpoints (and smooth) 
a single profile in \cite{Harchaoui}; there, they transformed the problem into a standard Lasso \cite{Tibshirani1996} and proposed
an extremely fast way to find the whole solution path, that is, an ordered set of pertinent breakpoints. When several profiles must be segmented simultaneously, we propose to impose a fused constraint jointly on the set of profiles, leading to a reformulation of the problem as  a group LARS or group Lasso \cite{Yuan}.
The  originality of our approach lies in the fact that the fused constraint 
imposes that each newly-selected breakpoint be shared by \emph{all} profiles. 
In this formulation, we end up  with a stupendously large matrix (with $p^2n^2$ entries) to work with, but we show that to run a group LARS
algorithm it is not necessary to store this matrix. 

In order to find $k$ breakpoints in $n$ profiles of $p$ probes, our method has a complexity $O(knp)$ in time and requires $O(np)$ in memory. In speed trials using Matlab on a 2008 Macbook Pro with 4GB of RAM, for $n=20$ profiles of $p=2000$  probes,  50 breakpoints were found in 0.72 seconds, an average of 0.014 seconds each. 
Even at the limit of current aCGH technology, with  $p\sim 946\ 000$ and, for example $n=20$, the first 50 selected breakpoints 
were obtained in 304 seconds,
an average of about 6 seconds each.
We present the performance of the algorithm on simulated data under various noise conditions, and demonstrate that it is able to recover the breakpoints shared by some or all of the profiles as the number of profiles increases. Finally we implement the algorithm to segment
aCGH profiles of bladder cancer patients.

\section{Methods}

Suppose we have $n$ aCGH profiles each of length $p$. The $p$ probe values are calculated at identical locations for each
of the $n$ profiles.
For each profile, we want to find a piecewise constant representation, where the jumps between constant segments represent 
\emph{breakpoints}, that is, places where the copy number changes.  

\subsection{One profile, one chromosome}

\begin{sloppypar}
Our starting point is the following framework for finding breakpoints in  a one-dimensional piecewise-constant signal with
white noise, as introduced by 
\cite{Harchaoui}.  Let $Y = (y_1,\ldots,y_p)$ be the observed signal. Consider the following constrained optimization problem:
\begin{equation}\label{eq:smooth}
\min_{\beta \in \mathbb{R}^p} \| \beta - Y  \|_2^2 \qquad \text{subject to} \qquad 
\sum_{i=1}^{p}  | \beta_{i} - \beta_{i-1} | < \mu ,
\end{equation}
where $\mu$ is a fixed non-negative constant and by convention $\beta_0 = 0$. The constraint $\sum_{i=1}^{p}  | \beta_{i} - \beta_{i-1} | < \mu$  
can be seen as a convex relaxation of bounding the number of jumps in $\beta$, an idea also implemented in the fused Lasso \cite{Tibshirani2008}. When $\mu$ is small enough, this constraint causes the solution $\beta$ to be made up of runs of equally-valued $\beta_i$ separated by an occasional jump from one constant to another, i.e., a piecewise constant function. 
For 
$\mu$ large enough the constraint is no longer effective and the solution is merely $\beta = Y$.
\end{sloppypar}

This  problem is convex, meaning that any standard convex optimization package can solve it for a given $\mu$ or a sequence 
$\{\mu_j\}_{j=1}^J$, though this remains computationally intensive. 
Making the change of variable $u_1 = \beta_1, u_2 = \beta_2 - \beta_1, \ldots, u_p = \beta_p - \beta_{p-1}$, \cite{Harchaoui} rewrite (\ref{eq:smooth}) as
$$
\min_{u \in \mathbb{R}^p}\| Au - Y  \|_2^2 \qquad \text{subject to} \qquad 
\sum_{i=1}^{p}  | u_p | < \mu ,
$$
where $A$ is a $p \times p$ matrix whose lower-diagonal and diagonal are $1$ and upper-diagonal is $0$. 
This is exactly a Lasso regression problem \cite{Tibshirani1996},  whose complete solution path for $\mu$ can be obtained for example by the {\it lars} package \cite{Efron} in a matter of seconds if $p$ is of the order of several hundred. However, once $p$ gets into the thousands, implementations like {\it lars}, which require  storing a $p\times p$ matrix and that have complexity $O(k^3 + p k^2)$ for calculating 
the first $k$ breakpoints, greatly slow down or do not run. It was shown in \cite{Harchaoui} that the  algorithm can be reformulated to take advantage of the structure of the matrix $A$ without storing it, resulting in a computation time in $O(pk)$ with only the need to store vectors of length $p$.

\subsection{Many profiles, one chromosome}

Let us now consider a $n \times p$ matrix $Y$ containing $n$ profiles of length $p$. Our aim is to apply a similar procedure to the one profile case,
but jointly to the whole set of 
profiles. We propose the following constrained optimization problem:
\begin{equation}\label{eq:smoothmulti}
\min_{\beta \in \mathbb{R}^{n \times p}} \| \beta - Y  \|_2^2 \qquad \text{subject to} \qquad 
\sum_{i=1}^{p}  \| \beta_{i} - \beta_{i-1} \|_2 < \mu ,
\end{equation}
where $\mu$ is a fixed non-negative constant, $\beta_i$ is the vector of length $n$ containing the value of the $i^{\text{th}}$ probe for
each of the $n$ individuals and by convention, $\beta_0 = \mathbf{0}$.
The choice of $L_2$ norm in the constraint has the effect of
creating long runs of consecutive vectors $\beta_i$ that are equal, interspersed with occasional jumps from one ``constant" vector to
the next \cite{Yuan}. This mirrors the one profile case with its piecewise constant approximation, but here it is $n$ profiles jointly.  
Intuitively, places where the set of profiles jointly ``break'' tend to be places where a large number of  individual profiles exhibit a break.

We now introduce a practical framework for solving such a problem. As for the one profile case \cite{Harchaoui}, we make the change
of variable $u_1 = \beta_1, u_2 = \beta_2 - \beta_1, \ldots, u_p = \beta_p - \beta_{p-1}$, where all these objects are now $n$-dimensional vectors. 
This gives us the representation:
$$\min_{u \in \mathbb{R}^{np}}\| Au - Y  \|_2^2 \qquad \text{subject to} \qquad 
\sum_{i=1}^{p}  \| u_i \|_2 < \mu ,$$
where $u$ is the $np$-dimensional vector of the $p$ vectors $u_i$ of length $n$ stacked on top of each other, by momentary abuse of notation $Y$ is the
$np$-dimensional vector of the columns of $Y$ stacked on top of each other and $A$ is now as follows: for each $1$ in the matrix $A$ of the
one profile case, replace it with an $n \times n$ identity matrix and for each $0$, replace it with an $n \times n$ matrix of zeros. The matrix $A$ thus becomes an
$np \times np$ matrix. 

In fact, this can be rewritten as a group Lasso \cite{Yuan}, i.e.,
\begin{equation}\label{grouplasso}
\min_{u \in \mathbb{R}^{np}}\left\| \sum_{i=1}^p A_iu_i - Y  \right\|_2^2 \qquad \text{subject to} \qquad 
\sum_{i=1}^{p}  \| u_i \|_2 < \mu ,
\end{equation}
where $A_i$ is the  matrix of size $np \times n$ of the columns $n(i-1) + 1$ up to $ni$ of $A$,
and $u_i$ the $i^{\text{th}}$ column of $u$.
 Here, each group $i$ is the set of $n$
variables, one from each profile, found in position $i$ on the genome.

The group Lasso and group LARS algorithms \cite{Yuan} select variables ``in groups'' rather than one by one as for Lasso and LARS. This is useful when we have prior knowledge as to relationships between subsets of variables. In our case, the relationship
is that each group represents the changes in value of $n$ profiles between two adjacent locations on the genome. Selection of a group by the
algorithm corresponds to choosing a location on the genome where a significant amount of change happens to a significant number of
profiles. Whereas the standard algorithms for solving Lasso and LARS are almost identical, generalizations to group Lasso and group LARS are less so. 
The group LARS algorithm we describe here is therefore not the solution path to (\ref{grouplasso}), but to a similar problem.
We chose to work with group LARS
as similar ideas to those in \cite{Harchaoui}  could be used to implement an extremely fast algorithm.

\subsection{``Many profiles, one chromosome'' fast group LARS \label{groupLARS}}

The group LARS algorithm we propose explicitly follows the steps given in  \cite{Yuan}, but avoids the suggested matrix formulation
by generalizing the methodology of \cite{Harchaoui}. To find breakpoints, we follow the following steps, which have 
computational complexity $O(np)$, resulting in a complexity $O(npk)$ to find the first $k$ breakpoints.

\begin{itemize}

\item \emph{Step 1:} start with  $u_{n \times p} = 0$, $k = 1$ and $r_{n \times p} = Y$. 

\begin{sloppypar}
\item \emph{Step 2:} compute the currently most correlated set:
$$
\mathcal{A}_1 = \arg\max_i \|A'_i r   \|_2^2 /p_i .
$$
By abuse of notation, $r$ has been momentarily written as an $np$-dimensional vector of the columns of $r$ stacked on top of each other. 
As $\forall i,j, \,p_i = p_j = n $ here, we can ignore the division by $p_i$. Due to the structure of $A$, to calculate the most correlated
set it suffices to calculate the
cumulative sums by row of $r_{n \times p}$, starting in the $p^{\text{th}}$ column and moving back along the rows to the $1^{\text{st}}$ column. We call this resulting $n \times p$
matrix $C$. The 
column of $C$ with the 
largest $L_2$ norm corresponds to the first selected group, i.e., breakpoint. This step takes $O(np)$ operations, and stores a $n\times p$ matrix.
\end{sloppypar}

\item \emph{Step 3:} compute the current direction $\gamma$. We represent $\gamma$ as an $n \times p$
matrix. Each column of $\gamma$ whose index is not in the active set $\mathcal{A}_k$ is identically zero.  Calculating the rest of $\gamma$ usually requires having to 
calculate $G_{\mathcal{A}_k}^{nk} = (A'_{\mathcal{A}_k}A_{\mathcal{A}_k})^{-1}$, the inverse of an $nk \times nk$ matrix. 
In fact, due to the analogous structure of $A$ with $A$ from the one profile case in \cite{Harchaoui}, it suffices to 
calculate the inverse $G_{\mathcal{A}_k}  =  G_{\mathcal{A}_k}^k$ of  the relevant $k \times k$ matrix  in the one profile case, which is a tridiagonal matrix with a closed form expression \cite{Harchaoui}.
%
Then, with the generic notation $B[:,i]$ meaning the $i^{\text{th}}$ column(s) of any matrix $B$, the remaining non-zero columns of $\gamma$ are obtained in $O(nk)$ by the matrix multiplication:
$$
\gamma[:,\mathcal{A}_k] \,\, = \,\, C[:,\mathcal{A}_k] \times G_{\mathcal{A}_k}'.
$$

\item \emph{Step 4:} for all $i \notin \mathcal{A}_k$, compute how far the group LARS algorithm will progress in direction $\gamma$ before
$A_i$ enters the most correlated set. This can be measured by an $\alpha_i \in [0,1]$ such that
$$
\|A'_i(r - \alpha_i A \gamma ) \|_2^2 \,\, = \,\, \|A'_{i'}(r - \alpha_i A \gamma ) \|_2^2,
$$
where $i'$ is arbitrarily chosen from $\mathcal{A}_k$ and $r$ and $\gamma$ are again momentarily given as $np$-dimensional vectors for notational simplicity. In \cite{Yuan}, it was shown that $\alpha_i$ is always well-defined.
This equation is quadratic in $\alpha_i$. To solve it simultaneously for all $i$ in $O(np)$:

\begin{enumerate}

\item define $\gamma^*$ as the $n \times p$ matrix we get by taking the row-wise cumulative sums of $\gamma$, starting this time at column 1 and
finishing at column $p$.

\item define $\delta^*$ as the $n \times p$ matrix we get by taking the row-wise cumulative sums of $\gamma^*$, starting  at column $p$ and
finishing at column 1.

\item the set of $p$ quadratic equations is symbolically $a\alpha^2 + b\alpha + c$, with $\alpha = (\alpha_1,\ldots,\alpha_p)$. Let 
$\mathcal{A}_k(j)$ mean the $j^{\text{th}}$ index in the active set, when sorted into ascending order. Then:

\item to calculate $a$, first let
$
a^* = \delta^* .\times \delta^* - (\delta^*[:,\mathcal{A}_k(1)] .\times \delta^*[:,\mathcal{A}_k(1)] ) \times \mathbf{1}_{1 \times p} , 
$
where `$\times$' means matrix multiplication and `$.\times$' means element by element multiplication.
$a$ is then the vector of length $p$ given by the column-wise sums of $a^*$.

\item to calculate $b$, let
$
b^* = C .\times \delta^* - (C[:,\mathcal{A}_k(1)] .\times \delta^*[:,\mathcal{A}_k(1)] ) \times \mathbf{1}_{1 \times p} . 
$
Then, $b$ is $-2$ times the column-wise sums of $b^*$.

\item to calculate $c$, let 
$
c^* = C .\times C - (C[:,\mathcal{A}_k(1)] .\times C[:,\mathcal{A}_k(1)] ) \times \mathbf{1}_{1 \times p} . 
$
Then, $c$ is the column-wise sums of $c^*$.

\item the quadratic equation can then be solved.

\end{enumerate}

\item \emph{Step 5:} select the smallest non-negative $\alpha_j$ such that $j \notin \mathcal{A}_k$, then update the active set to include this
new member:
$\mathcal{A}_{k+1} = \mathcal{A}_k \cup j$.

\item \emph{Step 6:} perform in $O(np)$ the updates: $u = u + \alpha_j \gamma$, $r = Y - A \gamma$ (with $\gamma$ momentarily written as an $np$-dimensional vector),
$k = k + 1$ and update $C$ with respect to the new $r$  as was done in Step 2. Then go back to Step 3 and repeat until either $p$
iterations are performed or a chosen number $K_{max} < p$ of breakpoints are found.

\end{itemize}

\subsection{Many profiles, many chromosomes}
CGH profiles usually span many chromosomes. Constraints linking the last data point on each chromosome to the first data point on the next are meaningless. 
Removing these constraints means that the matrix $A$ must be changed in the following way: replace 
$1$s with $0$s whenever they corresponds to entries of the matrix indexed by two different chromosomes. 
This has minor follow-on effects in the algorithm, including calculation of cumulative sums and the  matrix $G_{\mathcal{A}_k}$, and the
updating of $r$. The speed of the algorithm is unaffected.



\section{Experiments}


\subsection{Speed trials}\label{sec:speedtrial}

\begin{figure}
\begin{center}
\includegraphics[width=5.5in,height=2.5in]{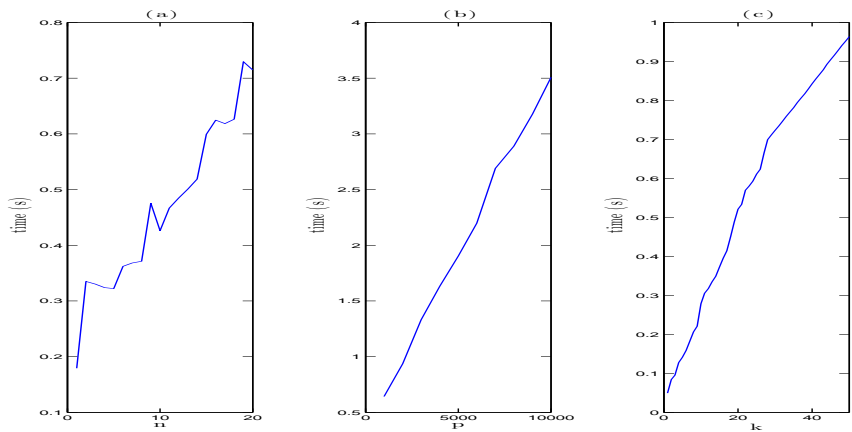} 
\end{center}
\caption{\textbf{Speed trials.}  \footnotesize{(a) CPU time
for finding 50 breakpoints when there are $2000$ probes and the number of profiles varies from $1$ to $20$.
(b)  CPU time when finding $50$ breakpoints with the number of profiles fixed at $20$ and the number of probes varying from $1000$ to $10000$ in intervals of $1000$.
(c)  CPU time for $20$ profiles and $2000$ probes
when selecting from $1$ to $50$ breakpoints.}
\label{Fig1}}
\end{figure}

All trials used Matlab on a 2008 Macbook Pro with 4GB of RAM.  Fig. \ref{Fig1}(a) indicates linearity in $n$, and
50 breakpoints were found in 0.72 seconds, an average of 0.014 seconds each.  
Fig. \ref{Fig1}(b) shows linearity in $p$.  Fig. \ref{Fig1}(c) shows, for $n$ and $p$ fixed, a near-linear relationship in $k$, i.e., 
subsequent breakpoints do not take longer to find than earlier ones. This confirms the theoretical $O(npk)$ complexity.

%

%
%


\begin{sloppypar}
To test the speed of the algorithm at the current limit of aCGH technology, we performed speed trials using sample data freely provided by
Affymetrix (http://www.affymetrix.com) for the Affymetrix Genome-Wide Human SNP Array 6.0, which includes around $946\ 000$ 
copy number probes. 
The  dataset includes 5 sets of 5 replicates. Three of the five individuals have abnormal copies of the X chromosome, with 3, 4 and 5 copies
respectively.  We calculated the  $\log$-ratio of 20 profiles (from 5 replicates on 4 individuals) against the first profile of the first (normal) individual after removing Y chromosome probes, leaving $937\ 223$ probes per profile. 
\end{sloppypar}


%
Again, the algorithm ran linearly in $k$. 
For $10$ profiles, the group LARS algorithm took $3.4$ seconds to find
each subsequent joint breakpoint, and for $20$ profiles, $6.1$ seconds.  Thus, the algorithm is computationally practical, extremely fast even, at the
upper limits of current technology.
We remark that for these $20$  profiles, the fast group LARS algorithm correctly selected as the most important joint breakpoint (out of the
$937\ 223$ possible choices) the jump from $0$ to the first probe on the X chromosome.  


\subsection{Performance on simulated data}

We performed a series of simulations in order to verify that the algorithm behaved well with respect to our stated goals, namely, recover breakpoints shared by several of a set of profiles. 
We designed four experiments  to move gradually from an artificial to a more realistic setting.

\begin{itemize}

\item[1.] 
\begin{sloppypar}
\emph{All profiles share the same breakpoints.} We simulated profiles of length $1000$, each divided into ten 
constant-valued segments of length 100. Hence, there are 10 breakpoints, located before probes 
$1, 101, 201,\ldots,901$. The constant value of each of the 10 segments was randomly drawn from a uniform
distribution on $[-1,1]$. White noise from a
$\mathcal{N}(0,\sigma^2)$ was then added independently to each of the 1000 probe values. As shown in Fig. \ref{simulation1},
we simulated with varying levels of noise: $\sigma^2 \in \{0.01, 0.1, 0.2, 0.5\}$. In particular, we see that with $\sigma^2 = 0.5$,
the noise is often significantly larger than the distance between subsequent underlying constant segments.
For a given value of $\sigma^2$, we randomly generated one profile in this way. We then asked the fast group LARS algorithm to find  
10 breakpoints only. If these ten breakpoints corresponded \emph{exactly} to the ten real breakpoints, we stopped. Otherwise, we added a
second randomly generated profile and ran the group LARS jointly on these two, and so on. 
For each of $1000$ such trials, we calculated the number of
profiles needed to find the 10 real breakpoints as its first 10 predictions.
A good algorithm should correctly select these breakpoints, given enough profiles.

\end{sloppypar}

\begin{figure}
\begin{center}
\includegraphics[width=3in,height=3.5in]{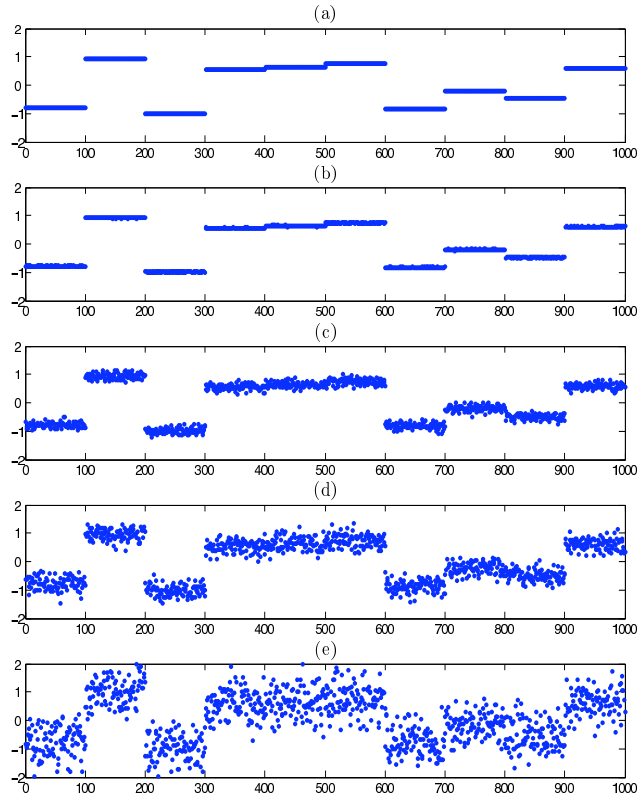} 
\end{center}
\caption{\textbf{Simulating an aCGH profile with different levels of white noise.}
\footnotesize{(a) is the underlying piecewise-constant profile. White noise from a $\mathcal{N}(0,\sigma^2)$ is added to the probes of (a)  with:
(b) $\sigma^2 = 0.01$, (c) $\sigma^2 = 0.1$, (d) $\sigma^2 = 0.2$ and (e) $\sigma^2 = 0.5$.}
\label{simulation1}}
\end{figure}


\item[2.] 
\emph{All profiles share the same breakpoint `regions', though the breakpoints are not all located at exactly the same probe on each profile.}
This is the same experimental condition as (1) except that each breakpoint that was fixed at location $i$ can now be at any of
$\{i-2, i-1, i, i+1, i+2\}$, chosen uniformly. 
For simplicity, the breakpoint before probe $1$ was kept fixed. 
For a given $\sigma^2$, we then run the algorithm as in (1), though in each loop we initially find the first 50 ordered breakpoints. 
We stop adding profiles one by
one when the following event happens: every one of the first $m \geq 10$ ordered predicted breakpoints is included in one of the $10$ breakpoint zones
and \emph{there is at least one predicted breakpoint in each of the $10$ zones}.
i.e., we find all of the $10$ zones before adding a non-existant breakpoint.

\item[3.] \emph{All profiles have a subset of a predefined set of breakpoints.} Potential breakpoints are 
located in the $10$ locations described in (1). First, the breakpoint before the first probe is automatically generated, i.e., the value of
the constant segment from probe $1$ to $100$. Then,
the potential breakpoint 2 before probe $101$ is randomly included  with probability $0.7$. If it is included, we randomly select
the value of the segment from $101$ to $200$ as in (1). Otherwise, the value of the segment from $1$ to $100$ is continued from $101$ to $200$.
We iterate this method up to the $10^{\text{th}}$ possible breakpoint.
Thus, on average there are $7.3$ breakpoints per profile (the first is always chosen and the $9$ others chosen independently with probability $0.7$). 
This is closer to what we see in reality, with aCGH profiles of patients with the same disease state sharing certain key breakpoints, yet 
but not all.

We then proceed as in (1) for $1000$ trials. One minor detail: if the first few randomly generated profiles only exhibit a subset of size
$s < 10$ of the $10$ possible breakpoints and the group LARS algorithm finds these $s$ breakpoints before any others, we stop the trial at this point,
as we cannot expect all $10$ breakpoints to be found if some of them have not yet been randomly exhibited.

\item[4.] \emph{All profiles have a subset of a predefined set of breakpoints though the exact location of each breakpoint 
can vary slightly between profiles.}
This experiment is a direct combination of (2) and (3). Again, $1000$ trials were performed for each level of noise. 
The minor detail mentioned in (3) is treated in the same way here.

\end{itemize}

\begin{figure}
\begin{center}
\includegraphics[width=2.5in,height = 3in]{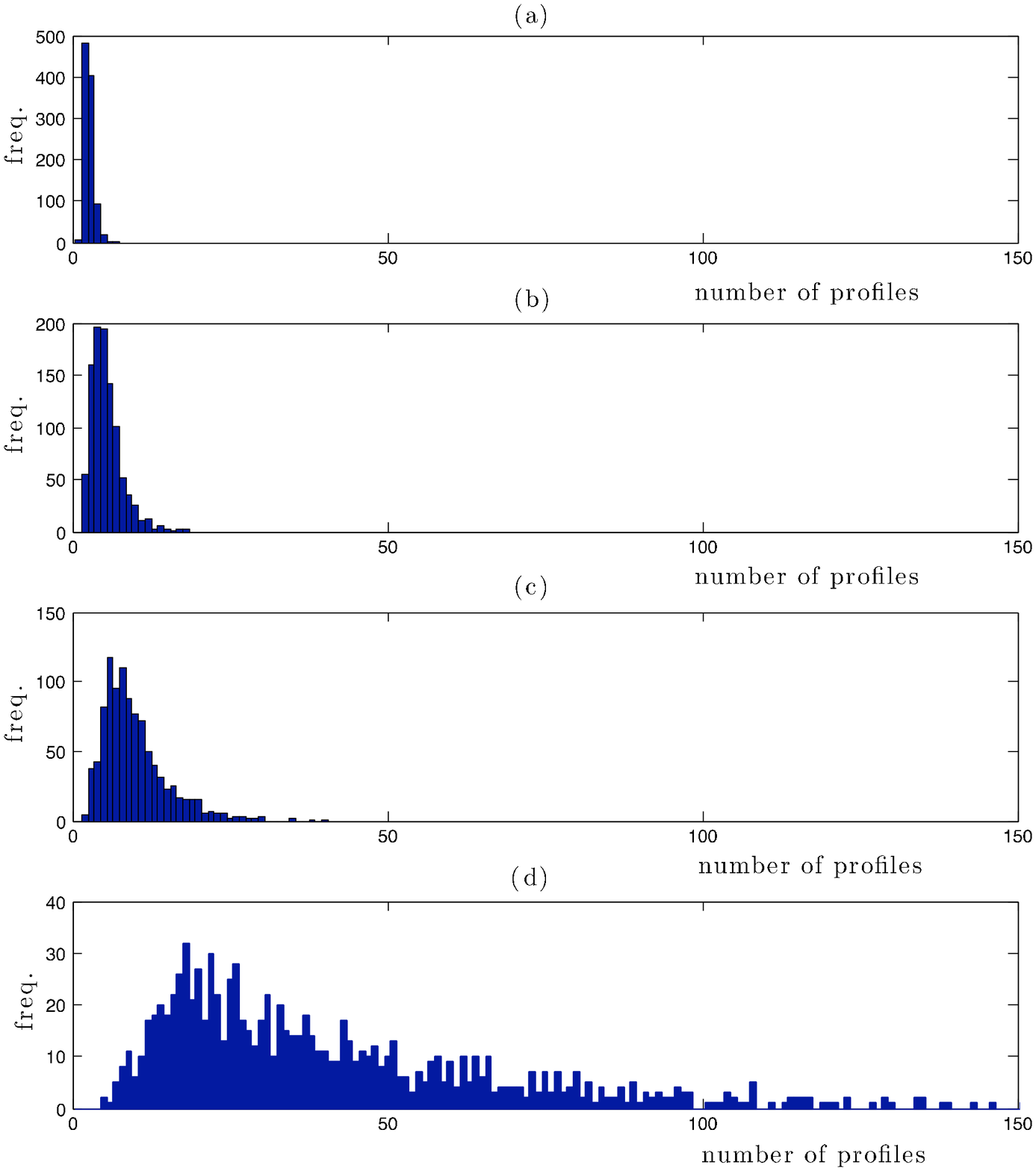} 
\includegraphics[width=2.5in,height=3in]{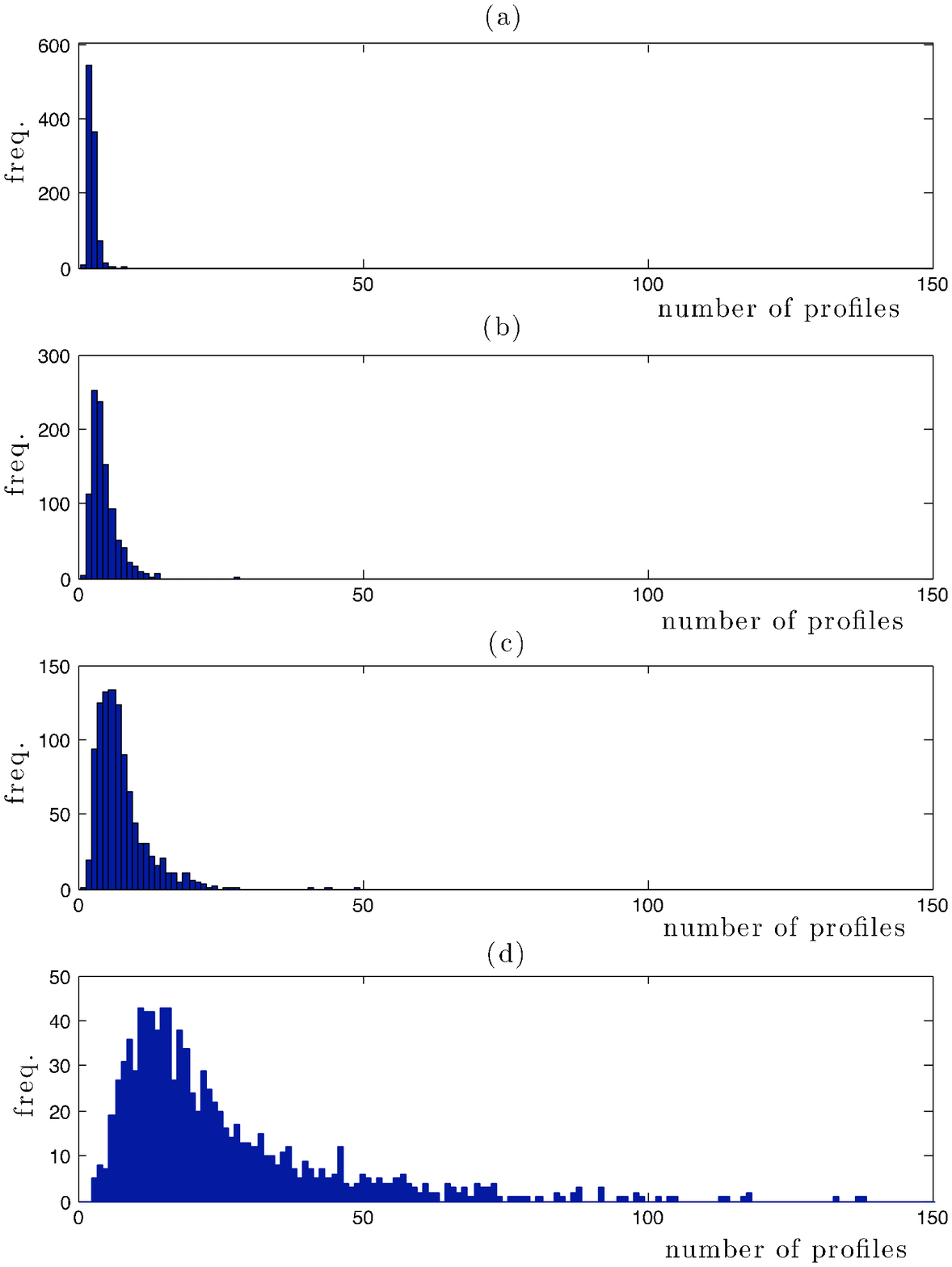} 
\end{center}
\caption{\textbf{Simulation conditions 1 (left) and 2 (right).}
\footnotesize{Left: histograms of the number of profiles required to correctly predict 10 real breakpoints with no mistakes in the presence of
white noise. Right: histograms of the number of profiles required to correctly predict all real breakpoints when each profile exhibits
a breakpoint in each of 10 tightly defined regions, in the presence of
white noise.
The noise is $\mathcal{N}(0,\sigma^2)$ with (a) $\sigma^2 = 0.01$, (b) $\sigma^2 = 0.1$, (c) $\sigma^2 = 0.2$ and
(d) $\sigma^2 = 0.5$. Each experiment was performed 1000 times.}
\label{hist1}}
\end{figure}

\begin{figure}
\begin{center}
\includegraphics[width=2.4in,height=3.1in]{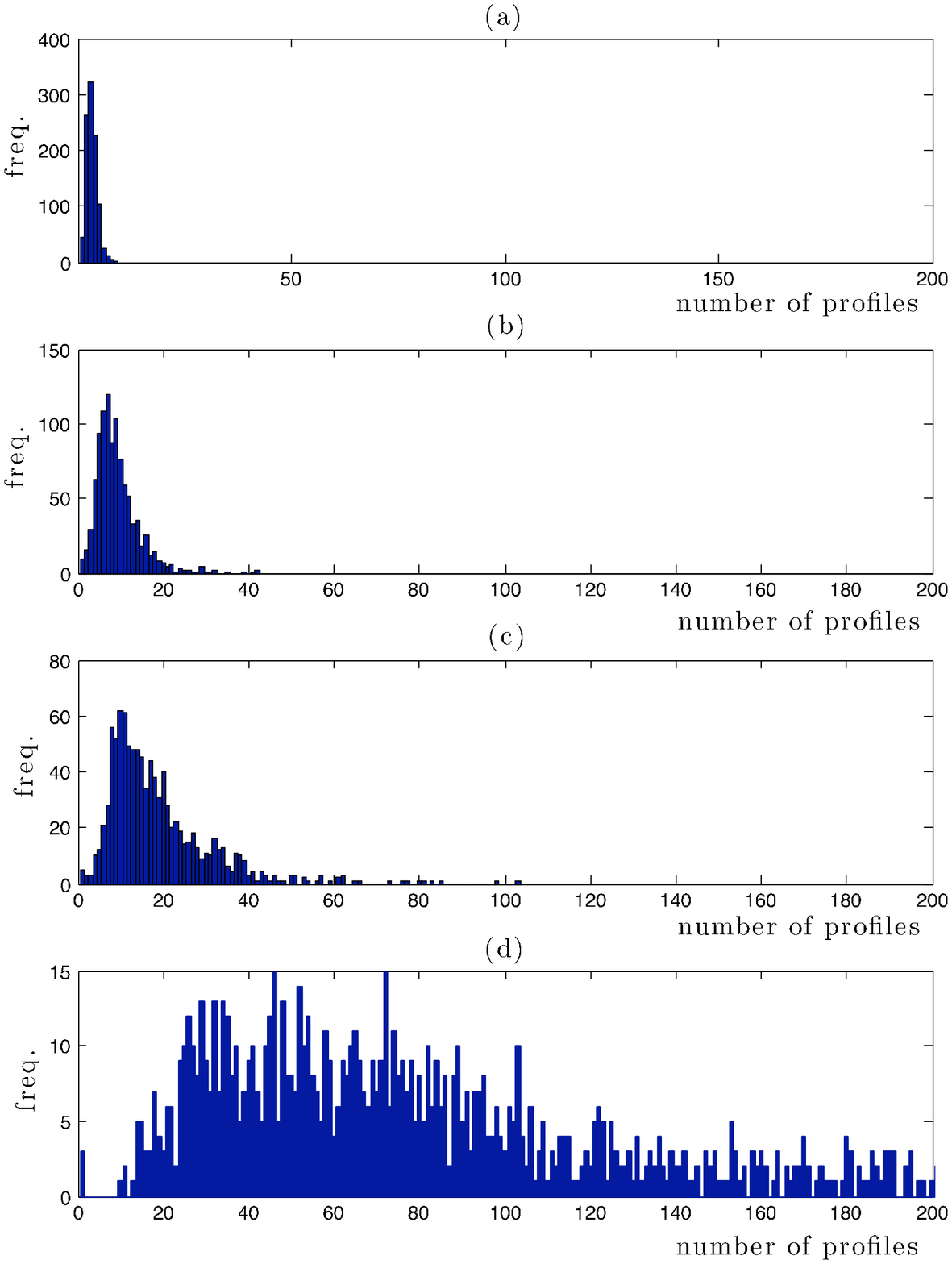} 
\hspace{.5cm}
\includegraphics[width=2.4in,height=3.1in]{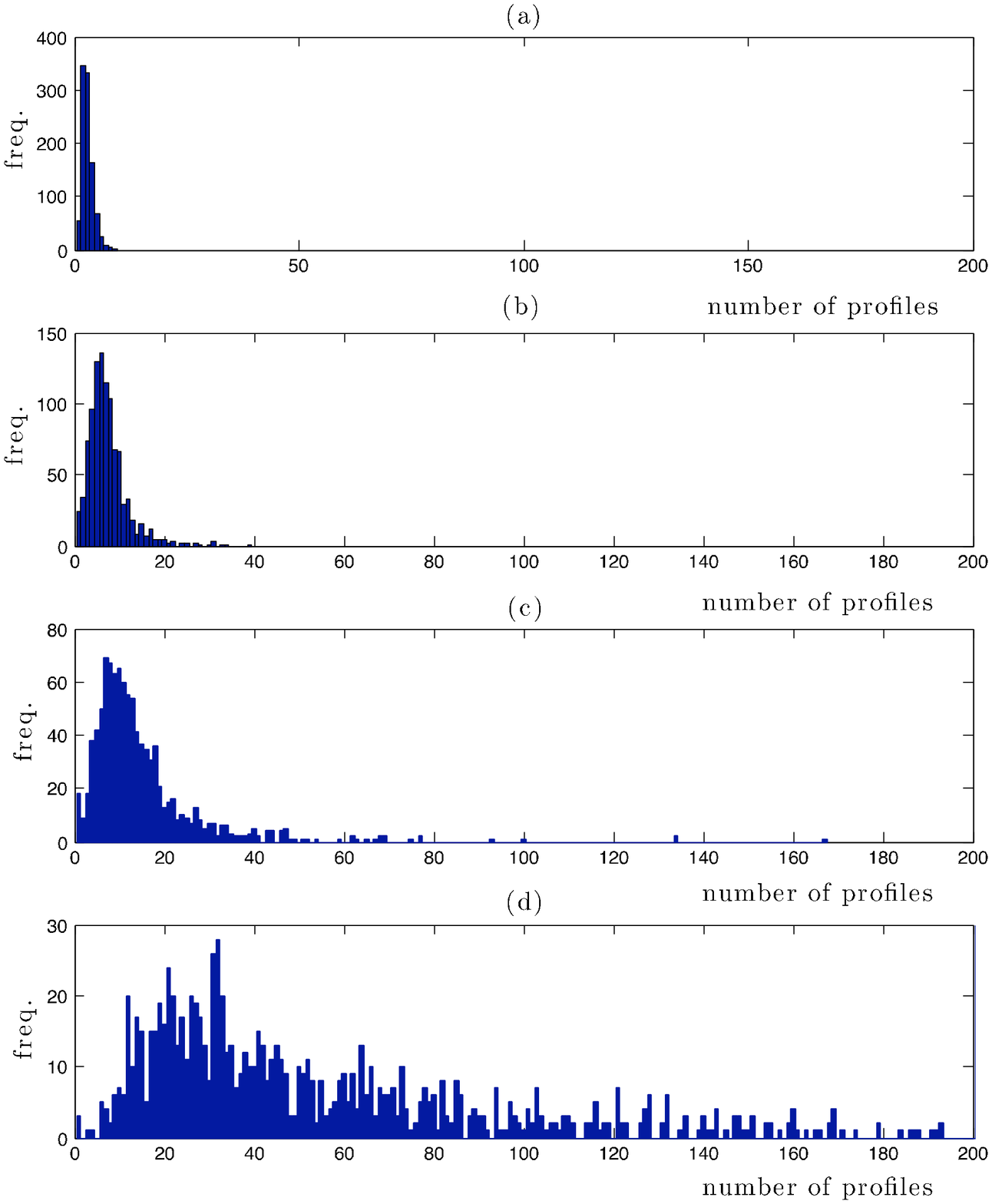} 
\end{center}
\caption{\textbf{Simulation conditions 3 (left) and 4 (right).}
\footnotesize{Left: histograms of the number of profiles required to correctly predict all real breakpoints when each profile exhibits
a subset of a predefined set of 10 breakpoints, in the presence of
white noise. Right: histograms of the number of profiles required to correctly predict all real breakpoints when each profile
exhibits a breakpoint in a subset of a predefined set of 10 tightly defined regions, in the presence of white noise.
The noise is $\mathcal{N}(0,\sigma^2)$ with (a) $\sigma^2 = 0.01$, (b) $\sigma^2 = 0.1$, (c) $\sigma^2 = 0.2$ and
(d) $\sigma^2 = 0.5$. Each experiment was performed 1000 times.}
\label{hist3}}
\end{figure}

Simulation results from experiments $1$-$4$ are shown in Figures \ref{hist1}-\ref{hist3}.
The main result is that, given enough profiles, the algorithm correctly selected the 10 breakpoint locations/regions for 
every experimental condition and  every noise level. Specifically,
in lower noise conditions
%
%
($\sigma^2 = 0.01$, $0.1$ and $0.2$), rarely are more than fifty profiles needed to correctly select the
true breakpoints. In more realistic conditions ($\sigma^2 = 0.5$), with high probability,  up to $75$-$200$ profiles were necessary to correctly select the 
whole set of true breakpoints, though often much fewer were required.

\subsection{Application to bladder tumor CGH profiles}

We considered a publicly available aCGH data set of 57 bladder 
tumor samples \cite{Stransky}. Each aCGH profile gave the 
relative quantity of DNA for 2215 probes. We removed the 
probes corresponding to sexual chromosomes, because the sex 
mismatch between some patients and the reference used made the 
computation of copy number less reliable, giving us a final list of 
2143 probes. 

Fig. \ref{piecewise}(a) shows the result of superimposing the smoothed versions of the 
$57$ bladder tumor aCGH profiles, when the algorithm has selected 80 ranked common breakpoints.
%
%
%
%
Figs \ref{piecewise}(b) and (c) show 2 of the original 57 profiles and their associated
smoothed version, where (b) was a profile exhibiting much instability, and (c) only on chromosome 9. 
We remark that even though (c) was forced to have the same breakpoints as (b), this does not
translate into a poor smoothed version of (c), rather, the forced breakpoints are tiny jumps that can be ignored by biologists.  

Fig. \ref{piecewise}(a) confirms nearly all of the duplications and deletions associated with bladder cancer found in 
\cite{ Kallioniemi1,Kallioniemi2,Blaveri2005Bladder}: frequent duplication of 8q22-24, 17q21 and 20q is observed, and frequent
deletion of 8p22-23, 13q, 17p, 11p and all of chromosome 9. The two known duplications that could not be confirmed here
were 12q14-15 and 11q13. Fig. \ref{piecewise}(a) suggests other potentially important CNVs, including frequent duplication of
1q, 5p  and deletion of 4q and 10q.



%

















\section{Discussion}

Segmentation of a single aCGH profile into regions of constant copy number, separated by breakpoints, is a well-studied 
problem \cite{Olshen2004Circular,Hupe, Picard2005,Fridlyand,Wang2005method,Picard2007,Huang2007Robust,Tibshirani2008}.
Recently, attempts have been made to deal simultaneously with many profiles \cite{Diskin,Rouveirol,Klijn,Picard2007b,Robin}. 
In these methods, the search for important shared CNV regions  tends to
occur as the final step, either by choice of a level of significance \cite{Diskin} or in post-processing \cite{Rouveirol}. These methods do
not use the biological  prior information that different profiles are likely to share at least some breakpoints.   
Also, they may require the  choice of pertinent kernel windows \cite{Klijn}, lose information through
discretization \cite{Diskin,Rouveirol} or involve prohibitive computational complexity for large $p$ \cite{Picard2007b}. 

\begin{figure}
\begin{center}
\includegraphics[width=4.5in,height=3.6in]{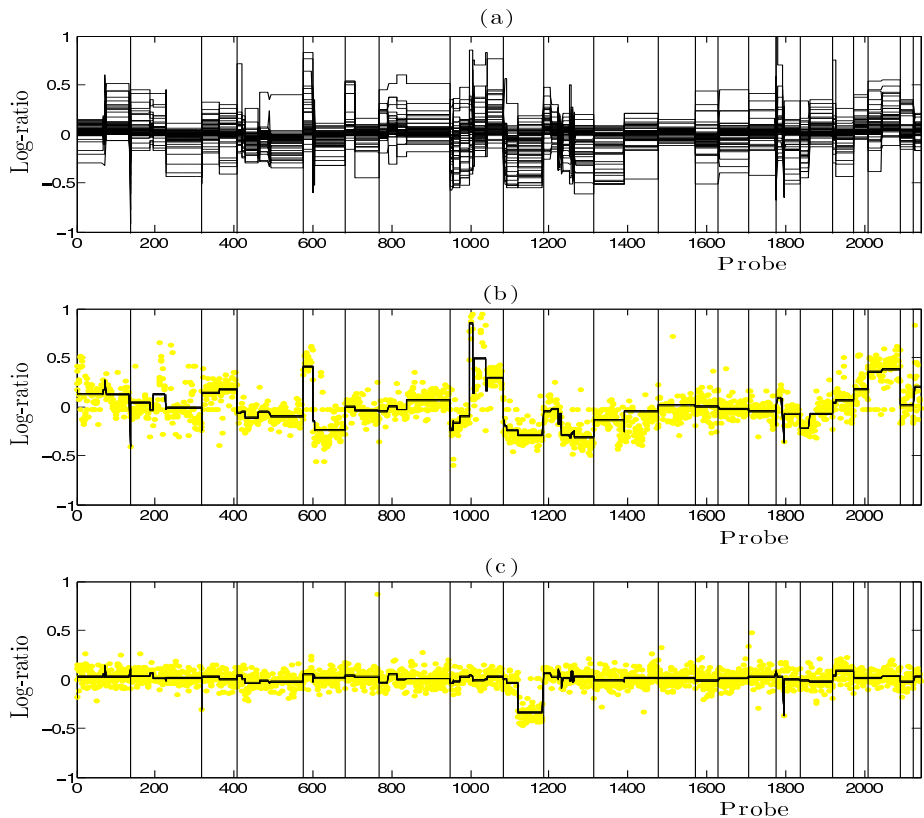} 
\end{center}
\caption{\textbf{Graphical representation.}
\footnotesize{(a) superimposition of the smoothed versions of
$57$ bladder tumor aCGH profiles \cite{Stransky} with 80 breakpoints. Vertical lines divide chromosomes 1-22.
(b) a profile exhibiting many CNVs, and its smoothed version. (c) a profile only showing a deletion on chromosome 9,
and its smoothed version. Smoothed profiles are obtained by replacing the set of probe values between consecutive breakpoints
with their mean value.}
\label{piecewise}}
\end{figure}

To our knowledge, we have introduced for the first time a way to explicitly code the prior biological information 
of expecting patients with the same disease to share 
certain CNVs. Our method forces breakpoints to be located in the same places for \emph{all} profiles. This has the
effect of
selecting breakpoint locations where many, but not necessarily all, profiles exhibit a breakpoint. This corresponds exactly to one of
the underlying biological goals in CNV studies. 
As shown in Fig. \ref{piecewise}(c), it is important to note that a profile forced 
to have breakpoints where it  clearly does not, still ends up with a good quality smoothed representation.
We also showed that
superimposing all smoothed versions on one graph allows intuitive visual interpretation of the data in a lower-dimensional form. 
On a real bladder cancer data set 
\cite{Stransky}, the smoothed versions we obtained (Fig. \ref{piecewise}(a)) confirmed nearly all of the known CNVs described in
the articles 
\cite{ Kallioniemi1,Kallioniemi2,Blaveri2005Bladder}.
Furthermore, our proposed algorithm is extremely fast. Even at the limit of current aCGH technology, it is practical, taking a few minutes 
on a single laptop computer.  

The piecewise-constant versions of the original profiles can be seen as extracted low-dimensional features. These can potentially
be used to implement classification algorithms to discriminate between two or more classes of aCGH profile, e.g., different disease
states. For example, each piecewise-constant profile could simply be treated as a vector of the constant values. As the 
breakpoints are forced to be in the same place on all profiles, the vector representation is the same size for each profile, directly
opening the way for the use of many well-known classification methods. This is a promising research direction.

The question of how many breakpoints to choose, i.e., when to `stop' the algorithm, remains open. There are at least two possible
solutions. First, if the algorithm were to be associated with a classification algorithm, a stopping criteria using internal cross-validation on the 
learning set can be defined. Second, it might be useful to calculate how much each
subsequently selected breakpoint closes the distance between the set of smoothed and original profiles, and define a 
stopping criteria based on this.  

\begin{thebibliography}{10}

\bibitem{Blaveri2005Bladder}
E.~Blaveri, J.~L. Brewer, R.~Roydasgupta, J.~Fridlyand, S.~DeVries, T.~Koppie,
  S.~Pejavar, K.~Mehta, P.~Carroll, J.~P. Simko, and F.~M. Waldman.
\newblock Bladder cancer stage and outcome by array-based comparative genomic
  hybridization.
\newblock {\em Clin Cancer Res}, 11(19 Pt 1):7012--7022, Oct 2005.

\bibitem{Bown}
N.~Bown, M.~Lastowska, S.~Cotterill, S.~O'Neill, C.~Ellershaw, P.~Roberts,
  I.~Lewis, and A.~D. Pearson.
\newblock 17q gain in neuroblastoma predicts adverse clinical outcome. {U}.{K}.
  cancer cytogenetics group and the {U}.{K}. children's cancer study group.
\newblock {\em Med. Pediatr. Oncol.}, 36:14--19, 2001.

\bibitem{Diskin}
S.~J. Diskin, T.~Eck, J.~Greshock, Y.~P. Mosse, T.~Naylor, C.~J.~Jr Stoeckert,
  B.~L. Weber, J.~M. Maris, and G.~R. Grant.
\newblock {STAC}: a method for testing the significance of {DNA} copy number
  aberrations across multiple array-{CGH} experiments.
\newblock {\em Genome Res.}, 16:1149--1158, 2006.

\bibitem{Efron}
B.~Efron, I.~Johnstone, T.~Hastie, and R.~Tibshirani.
\newblock {L}east {A}ngle {R}egression.
\newblock {\em Ann. Stat.}, 32(2):407--499, 2003.

\bibitem{Fridlyand}
J.~Fridlyand, A.~Snijders, D.~Pinkel, D.~Albertson, and A.~Jain.
\newblock Hidden markov models approach to the analysis of array {CGH} data.
\newblock {\em J. Multivariate Anal.}, 90:132--153, 2004.

\bibitem{Gershon}
D.~Gershon.
\newblock {DNA} microarrays: more than gene expression.
\newblock {\em Nature}, 437:1195--1198, 2005.

\bibitem{Harchaoui}
Z.~Harchaoui and C.~L\'evy-Leduc.
\newblock {C}atching change-points with lasso.
\newblock In {\em Adv. {N}eural {I}nform. {P}rocess. {S}yst. 22}, volume~22,
  2008.

\bibitem{Huang2007Robust}
Jian Huang, Arief Gusnanto, Kathleen O'Sullivan, Johan Staaf, Ake Borg, and
  Yudi Pawitan.
\newblock Robust smooth segmentation approach for array cgh data analysis.
\newblock {\em Bioinformatics}, 23(18):2463--2469, Sep 2007.

\bibitem{Hupe}
P.~Hup\'e, N.~Stransky, J.~P. Thiery, F.~Radvanyi, and E.~Barillot.
\newblock array {CGH} data: from signal ratio to gain and loss of {DNA}
  regions.
\newblock {\em Bioinformatics}, 20:3413--3422, 2004.

\bibitem{Kallioniemi2}
A.~Kallioniemi, O.~P. Kallioniemi, G.~Citro, G.~Sauter, S.~Devries,
  R.~Kerschmann, P.~Caroll, and F.~Waldman.
\newblock Identification of gains and losses of {DNA} sequences in primary
  bladder cancer by comparative genomic hybridization.
\newblock {\em Gene Chromosome Canc}, 12:213--219, 1995.

\bibitem{Kallioniemi1}
A.~Kallioniemi, O.~P. Kallioniemi, D.~Sudar, D.~Rutovitz, J.~W. Gray,
  F.~Waldman, and D.~Pinkel.
\newblock Comparative genomic hybridization for molecular cytogenetic analysis
  of solid tumors.
\newblock {\em Science}, 258:818--821, 1992.

\bibitem{Klijn}
C.~Klijn, H.~Holstege, J.~de~Ridder, X.~Liu, M.~Reinders, J.~Jonkers, and
  L.~Wessels.
\newblock Identification of cancer genes using a statistical framework for
  multiexperiment analysis of nondiscretized array {CGH} data.
\newblock {\em Nucleic Acids Res.}, 36(2):e13, 2008.

\bibitem{Olshen2004Circular}
A.~B. Olshen, E.~S. Venkatraman, R.~Lucito, and M.~Wigler.
\newblock Circular binary segmentation for the analysis of array-based {DNA}
  copy number data.
\newblock {\em Biostatistics}, 5(4):557--572, Oct 2004.

\bibitem{Picard2007b}
F.~Picard, \'E. Lebarbier, E.~Budinsk\'a, and S.~Robin.
\newblock Joint segmentation of multivariate {G}aussian processes using mixed
  linear models.
\newblock {\em Research Report}, 2007.

\bibitem{Picard2005}
F.~Picard, S.~Robin, M.~Lavielle, C.~Vaisse, and J.-J. Daudin.
\newblock A statistical approach for array {CGH} data analysis.
\newblock {\em BMC Bioinformatics}, 6:27, 2005.

\bibitem{Picard2007}
F.~Picard, S.~Robin, E.~Lebarbier, and J.-J. Daudin.
\newblock A segmentation-clustering problem for the analysis of array {CGH}
  data.
\newblock {\em Biometrics}, 63:758--766, 2007.

\bibitem{Pinkel}
D.~Pinkel, R.~Segraves, D.~Sudar, S.~Clark, I.~Poole, D.~Kowbel, C.~Collins,
  W.-L. Kuo, C.~Chen, Y.~Zhai, S.~H. Dairkee, B.-M. Ljung, J.~W. Gray, and
  D.~G. Albertson.
\newblock High resolution analysis of {DNA} copy number variation using
  comparative genomic hybridization to microarrays.
\newblock {\em Nat. Genet.}, 20:207--211, 1998.

\bibitem{Robin}
S.~Robin and V.~T. Stefanov.
\newblock Simultaneous occurrences of runs in independent {M}arkov chains.
\newblock {\em Meth. Comput. Appl. Probab.}, 11(2):267--275, 2008.

\bibitem{Rouveirol}
C.~Rouveirol, N.~Stransky, P.~Hup\'e, P.~La~Rosa, E.~Viara, E.~Barillot, and
  F.~Radvanyi.
\newblock Computation of recurrent minimal genomic alterations from array-{CGH}
  data.
\newblock {\em Bioinformatics}, 22(7):849--856, 2006.

\bibitem{Speicher}
M.~Speicher, G.~Prescher, S.~du~Manoir, A.~Jauch, B.~Horsthemke, N.~Bornfeld,
  R.~Becher, and T.~Cremer.
\newblock Chromosomal gains and losses in uveal melanomas detected by
  comparative genomic hybridization.
\newblock {\em Clin. Cancer Res.}, 11:7012--7022, 2005.

\bibitem{Stransky}
N.~Stransky, C.~Vallot, F.~Reyal, I.~Bernard-Pierrot, S.~Gil Diez~de Medina,
  R.~Segraves, Y.~de~Rycke, P.~Elvin, A.~Cassidy, C.~Spraggon, A.~Graham,
  J.~Southgate, B.~Asselain, Y.~Allory, C.~C. Abbou, D.~G. Albertson, J.-P.
  Thiery, D.~K. Chopin, D.~Pinkel, and F.~Radvanyi.
\newblock Regional copy number-independent deregulation of transcription in
  cancer.
\newblock {\em Nat. Genet.}, 38:1386--1396, 2006.

\bibitem{Tibshirani1996}
R.~Tibshirani.
\newblock Regression shrinkage and selection via the lasso.
\newblock {\em J. Royal. Statist. Soc. B.}, 58:267--288, 1996.

\bibitem{Tibshirani2008}
R.~Tibshirani and P.~Wang.
\newblock Spatial smoothing and hot spot detection for {CGH} data using the
  fused lasso.
\newblock {\em Biostatistics}, 9(1):18--29, 2008.

\bibitem{VanRoy}
N.~Van~Roy, J.~Vandesompele, G.~Berx, K.~Staes, M.~Van~Gele, E.~De~Smet,
  A.~De~Paepe, G.~Laureys, P.~van~der Drift, R.~Versteeg, F.~Van~Roy, and
  F.~Speleman.
\newblock Localization of the 17q breakpoint of a constitutional 1;17
  translocation in a patient with neuroblastoma within a 25-kb segment located
  between the accn1 and tlk2 genes and near the distal breakpoints of two
  microdeletions in neurofibromatosis type 1 patients.
\newblock {\em Gene Chromosome Canc}, 35:113--120, 2002.

\bibitem{Wang2005method}
P.~Wang, Y.~Kim, J.~Pollack, B.~Narasimhan, and R.~Tibshirani.
\newblock A method for calling gains and losses in array {CGH} data.
\newblock {\em Biostatistics}, 6(1):45--58, Jan 2005.

\bibitem{Yao}
J.~Yao, S.~Weremowicz, B.~Feng, R.~C. Gentleman, J.~R. Marks, R.~Gelman,
  C.~Brennan, and K.~Polyak.
\newblock Combined c{DNA} array comparative genomic hybridization and serial
  analysis of gene expression analysis of breast tumor progression.
\newblock {\em Cancer Res.}, 66:4065--4078, 2006.

\bibitem{Yuan}
M.~Yuan and Y.~Lin.
\newblock Model selection and estimation in regression with grouped variables.
\newblock {\em J. R. Statist. Soc. B}, 68:49--68, 2006.

\end{thebibliography}

\end{document}